\documentstyle[11pt,epsf]{article}

\def\reference{\noindent \hskip -0.5cm}
\def\apj{ApJ}
\def\astap{AA}
\def\mnras{MNRAS} 

\begin{document}
\begin{flushright}
\noindent{\tt astro-ph/9612020}\\
Invited talk at the {\it HST User's Meeting}\\
{\it Baltimore, Sept.12-13, 1996}\\
\end{flushright}

\begin{center}
{\Large\bf Deuterium abundance and cosmology}

\vskip 0.5cm

\noindent 
A. Vidal-Madjar$^1$, R. Ferlet$^1$ and M. Lemoine$^2$
\end{center}
\vskip 0.5cm

\noindent
1 Institut d'Astrophysique de Paris, CNRS, 98bis Bd. Arago, 75014 Paris, 
France\\
2 Department of Astronomy \& Astrophysics, Enrico Fermi 
Institute, The University of Chicago, Chicago, IL 60637-1433
\vskip 1.0cm

\noindent
{\leftskip 1cm
\rightskip 1cm

\noindent
{\small{\bf Abstract.} 
We review the status of the measurements of the deuterium abundance from the
local interstellar medium to the solar system and high redshifts absorbers 
toward quasars. We present preliminary results toward a white dwarf and a QSO.
We conclude that the deuterium evolution from the Big Bang to now is still not
properly understood.
}

}
\vskip 1.cm

\section{Introduction}

It is well accepted that deuterium is only produced in significant amount
during primordial nucleosynthesis (BBN, see e.g. the proceedings of the ESO 
Symposium {\it The Light Element Abundances}, 1995, Crane Ed., Springer). 
Moreover, D is thoroughly destroyed in stellar interiors. Hence, any abundance 
of deuterium measured at any metallicity should provide a lower limit to the 
primordial deuterium abundance. Deuterium is thus a key element in cosmology 
and in galactic chemical evolution (see e.g. Audouze \& Tinsley 1976;
Gautier \& Owen 1983; Vidal-Madjar \& Gry 1984; Boesgaard \& Steigman 1985; 
Olive et al. 1990; Pagel 1992; Vangioni-Flam \& Cass\'e 1994; Prantzos 1996).
Indeed, its primordial abundance is the best tracer of the baryonic density 
parameter of the Universe $\Omega_B$, and the decrease of its abundance along 
the galactic evolution should trace the amount of star formation (among others).

The first, although indirect, measurement of the deuterium abundance of 
astrophysical significance was carried out through $^3$He evaluation in the 
solar wind, leading to
D/H$\simeq2.5\pm1.0\times10^{-5}$ (Geiss \& Reeves 1972), a value 
representative of 4.5 Gyrs ago. The first measurements of the interstellar D/H 
ratio, representative of the present epoch, were reported shortly thereafter 
(Rogerson \& York 1973). Their value of D/H$\simeq1.4\pm0.2\times10^{-5}$
has since then nearly not changed, whatever the availability of adequate 
instrumentation was. For more than a decade, these interstellar abundances 
have been used to constrain BBN in a direct way.

However, abundances measured at lower metallicities are less contaminated by 
the effect of galactic evolution. This is the reason why the deuterium 
abundance is now being chased in high redshift absorbers on quasar lines of 
sight: the composition of these clouds of very low metallicity should reflect 
the actual primordial deuterium value.

In the following, we discuss the different measurements of the deuterium 
abundance, following an anti-chronological order: in the ISM, in the pre-solar
nebula, and in high redshift absorbers. We conclude eventually that many 
aspects of the evolution of deuterium in the Universe are yet unknown.

\section{Interstellar observations}

There are several methods to measure the interstellar abundance of deuterium 
(see Vidal-Madjar 1991, Ferlet 1992). One of them is to observe deuterated 
molecules such as HD, DCN, {\it etc}... and to form the ratio of the 
deuterated molecule column density to its non-deuterated counterpart 
(H$_2$, HCN, {\it etc.}...). More than twenty different deuterated species 
have been identified in the ISM, with abundances relative to the non-deuterated 
counterpart ranging from $10^{-2}$ to 10$^{-6}$. Conversely, this means that 
fractionation effects are important, and that, as a consequence, this method 
cannot provide a precise estimate of the true interstellar D/H ratio; rather, 
this method is used in conjunction with estimates of the interstellar D/H 
ratio to gather information on the chemistry of the ISM.

Another way to derive the D/H ratio comes through radio observations of the 
hyperfine line of D{\sc i} at 92cm. The detection of this line is however
extremely difficult, and no firm detection has ever been reported. The 
detection of this line would allow to probe more distant interstellar media 
than the local medium discussed below; however, because a large column density 
of D is necessary to provide even a weak spin-flip transition, these 
observations aim at molecular complexes. As a result, the upper limit derived 
toward Cas A (Heiles et al. 1993): D/H$\leq2.1\times10^{-6}$ may as well 
result from a large differential fraction of D and H being in molecular form
in these clouds, as from the fact that one expects the D/H ratio to be lower
closer to the galactic center (since D is destroyed in stellar processing). 

Finally, the only way to derive a reliable estimate of the interstellar D/H 
ratio is to observe the atomic transitions of D and H of the Lyman series in 
the far-UV, in absorption in the local ISM against the background continuum of 
cool or hot stars. These observations have been performed using the Copernicus
and the IUE satellites, and now the Hubble Space Telescope. Both types of 
target stars present pros and cons.

\subsection{Cool stars}

The main advantage of observing cool stars is that they can be selected in the 
vicinity of the Sun. This results in low H{\sc i} column densities, and 
``trivial'' to nearly ``trivial'' lines of sight. In effect, due to the low 
atomic weight of H{\sc i} and D{\sc i}, to the D{\sc i}--H{\sc i} $-$82km/s 
isotopic shift, and to the abundance of H{\sc i} in the local medium, the 
D{\sc i} line cannot be detected at Lyman $\alpha$ in the wing of the H{\sc i} 
line for H{\sc i} column densities larger than $10^{19}$ cm$^{-2}$. Also, the 
presence of several interstellar components with different $b$-values may 
imply a large error on the H{\sc i} column density if these components are 
unresolved. For this reason, deriving the H{\sc i} column density has always 
been the limiting factor of accurate D/H ratios measurements. Note that the 
spectral resolutions of Copernicus and IUE were respectively 15 and 
30 km s$^{-1}$, and, as a consequence, a non-trivial line of sight, even in 
the local ISM, would generally go unresolved. Eventhough HST--GHRS now offers a 
spectral resolution of 3.5 km s$^{-1}$, the thermal width of the D{\sc i} line 
in the local ISM is $\simeq8$ km s$^{-1}$, so that one has to observe lines of 
heavier species (thinner lines) to fully use the resolving power of HST, and 
build up a coherent line of sight velocity structure. This is one of the first 
difficulties inherent to the ``cool stars'' approach: the detailed structure 
of the line of sight could be found only through the observation of the 
Fe{\sc ii} and the Mg{\sc ii} ions, which are unfortunately present in both
H{\sc i} and H{\sc ii} regions and thus may not trace properly the H{\sc i} 
gas. In particular, species like N{\sc i} and O{\sc i} could not be observed 
(see below).

Moreover, the chromospheric Lyman $\alpha$ emission line has to be modeled 
to set the continuum for the interstellar absorption. Such a procedure 
necessarily introduces systematic errors. 

Nevertheless, this method has provided the most precise measurement of the 
local D/H ratio in the direction of Capella, using HST--GHRS:
(D/H)$_{\alpha Aur}=1.60\pm0.09^{+0.05}_{-0.10}\times10^{-5}$ 
(Linsky et al. 1993, 1995), assuming no systematics. 

In that unique case, the possible systematics due to the chromospheric line 
profile are probably well reduced because the observations of that 
spectroscopic binary were made at two different phases of the system when the 
two stellar chromospheric lines were placed very differently in the velocity 
space owing to the stars motion within the system. The result is that very 
different and independent observations were consistent with only one 
interstellar absorption and two chromospheric line profiles simply shifted 
from one epoch to the other. Capella is also unique in the local ISM by 
revealing apparently a simple line of sight with a single component. If true, 
although this could always be debated (see below), the confidence in the 
column densities evaluation could be relatively high.

Several more cool stars have been observed with HST since then 
(Linsky et al. 1995: Capella, Procyon; 
Linsky \& Wood 1996: $\alpha$~Cen A, $\alpha$~Cen B; 
Piskunov et al. 1996: HR 1099, 31 Com, $\beta$~Cet, $\beta$~Cas; 
Dring et al. 1997: $\alpha$~Tri, $\epsilon$~Eri, $\sigma$~Gem, $\beta$~Gem). 
Although all compatible with the Capella evaluation, none of these results is 
precise enough to place any new constraints on this evaluation.

\subsection{Hot stars}

Hot stars are unfortunately located further away from the Sun, so that one 
always has to face a high H{\sc i} column density and often a non-trivial 
line of sight structure. In these cases, D{\sc i} could not be detected at 
Ly$\alpha$, and one has to observe higher order lines, {\it e.g.} Ly$\gamma$,
Ly$\delta$, Ly$\epsilon$; hence these measurements have primarily come through 
Copernicus observations. The stellar continuum is however smooth at the 
location of the interstellar absorption and, moreover, the N{\sc i} triplet 
at 1200~\AA ~as well as other N{\sc i} lines are available to probe the 
velocity structure of the line of sight. In particular, N{\sc i} and O{\sc i} 
were shown to be excellent tracers of H{\sc i} in the ISM (Ferlet 1981; 
York et al. 1983). The interstellar void identified in the direction to hot 
stars in the Canis Majoris tunnel (Gry et al. 1995) has allowed HST 
observations of D{\sc i} at Ly$\alpha$, but no further constraints have 
resulted so far.

\subsection{Present status}

All published D/H ratios are collected in fig. 1, distinguishing hot stars from 
cool stars observations (all references except Allen et al. 1992 can be found 
in Vidal-Madjar 1991, Ferlet 1992; note that the most recent cool stars
estimations from HST are not included). The D/H ratios range from 
$\sim5\times10^{-6}$ to $\sim4\times10^{-5}$. A large scatter is clearly 
detected in fig. 1 and represents differences of the D/H ratio in the local 
ISM, that may be as large as a factor $\simeq4$ over scales as small as a 
few parsecs. The essential question is: do these variations really exist?

\begin{figure}[h]
\epsfxsize=14cm
$$\epsfbox[20 50 540 310]{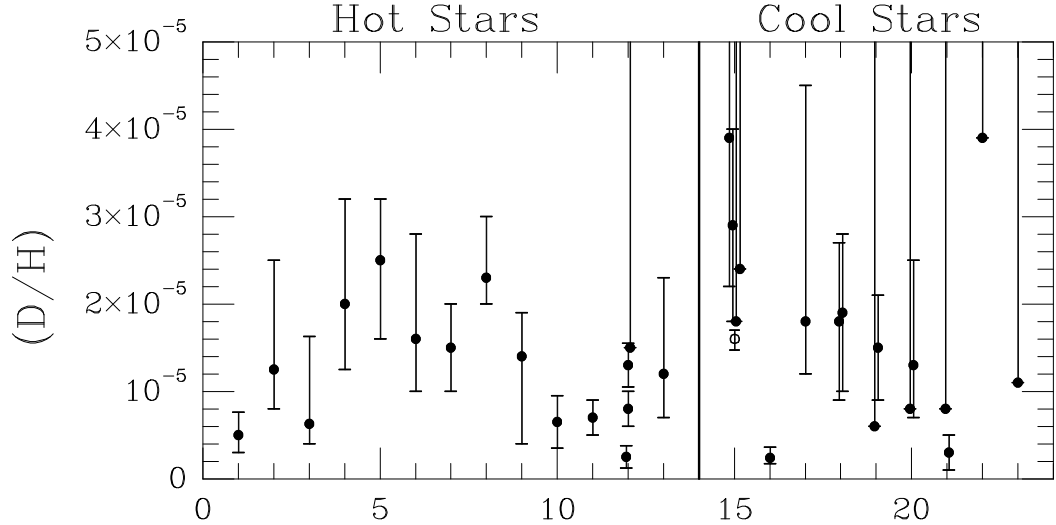}$$
\caption{Measurements of the D/H ratio in the local ISM. The left hand-side 
box collects data obtained toward hot stars, while the right hand-side one 
shows cool stars observations. The $x$-axis has no physical significance, and 
merely labels the different stars. Data points next to each other, within 
less than 1 $x$-axis unit, correspond to the same target star. The open circle 
represents the Linsky et al. (1993, 1995) measurement using HST toward Capella.
All other data points come from Copernicus or IUE observations.}
\label{dsh}
\end{figure}

Unfortunately, one cannot answer this question and at the same time be 
perfectly objective. On one hand, note that the Linsky et al. (1993, 1995) 
measurement does not agree with any of the previous D/H measurements toward 
Capella. However, one cannot ignore this measurement of unprecedented quality.
On the other hand, one could use the observed scatter between different 
measurements toward a same star to get a rough estimate of the systematics; 
such an estimate does not seem to be able to account for the large scatter of 
fig. 1 for all the other stars. Finally, it could be tempting -- but also very 
arbitrary -- to claim that the systematics associated with IUE and Copernicus 
observations are large, that they account for the observed discrepancies, and 
that only the Linsky et al. (1993, 1995) value should be kept. Although 
undoubtedly of great quality, the Capella measurement comes through the 
modeling of the chromospheric emission line of an unresolved binary system and 
the detection of only one absorbing component on the line of sight, through 
Mg{\sc ii} and Fe{\sc ii} observations. As mentioned above, it could well be 
that H{\sc i} media, shifted from the Local Cloud by a few km s$^{-1}$ went 
unnoticed in Mg{\sc ii} and Fe{\sc ii} although they would play an important 
role in the H{\sc i} saturated profile. Such systematics were not considered 
by Linsky et al. (1993, 1995).

To answer to the reliability of the observations shown in fig. 1, one has to 
re-analyze all these data in a consistent way, looking for possible undetected 
systematics. One has to recall as an example that time variations of the D/H 
ratio have already been reported toward $\epsilon$ Per (Gry et al. 1983), 
which were interpreted as due to the ejection of high velocity hydrogen atoms 
from the star. This perturbation can only enhance the D/H ratio. It is worth 
noting that in at least four cases the D/H ratio was found to be really low:
$0.7\pm0.2\times10^{-5}$ and $0.65\pm0.3\times10^{-5}$~toward $\delta$~and 
$\epsilon$~Ori (Laurent et al. 1979); $0.8\pm0.2\times10^{-5}$~toward 
$\lambda$~Sco (York 1983) and $0.5\pm0.3\times10^{-5}$~toward $\theta$~Car 
(Allen et al. 1992). In each cases, authors discussed in details possible 
systematics but concluded that none of the identified ones could explained 
these low nearby ISM D/H values.

Finally, note that various explanations to these possible fluctuations of the 
D/H ratio have been put forward as early as Vidal-Madjar et al. (1978),
Bruston et al. (1981) as a selective radiation pressure effect. More recently,
even the very possibility of measuring the D/H ratio was questionned as a
consequence of possible stochastic velocity fields with finite correlation
lengths in the interstellar medium (Levshakov et al. 1996). The consequence
could be that any evaluation made through line profile analysis assuming Voigt
profiles, can lead to very different evaluations of the D/H ratio for a unique 
assumed one. This has been discussed by Jenkins (1996) who shows that with a 
sample of lines presenting different oscillator stengths, it is possible to 
evaluate correctly the column densities whatever the velocity distribution of 
the cloud is. Such a situation is fortunately often the case in the Copernicus 
studies of the D/H ratio. This is however a long-standing problem. 

To try to make some progress, we have inaugurated the use of a new type of 
targets that should solve many of the intrinsic difficulties of the problem, 
namely nearby white dwarfs.

\subsection{Actual prospects}

Observing white dwarfs has many advantages. Such targets can be chosen near to
the Sun, circumventing the main disadvantage of hot stars, and they can also 
be chosen in the high temperature range, so as to provide a smooth stellar 
profile at Ly$\alpha$. At the same time, the N{\sc i} triplet at 1200\AA~as 
well as the O{\sc i} line at 1302\AA~would be available, allowing thus an 
accurate sampling of the line of sight. Such observations have now been 
conducted using HST toward two white dwarfs: G191--B2B (Lemoine et al. 1996; 
Vidal-Madjar et al. 1997) and Hz43 (Landsman et al. 1996).

In the case of Hz43, the structure of the line of sight appears trivial at a
first glance, {\it i.e.} it only consists of the Local Cloud. The D/H ratio as 
well as the H{\sc i} column density are consistent, in the single-cloud 
hypothesis, with those of the Local Cloud obtained by Linsky et al. 
(1993, 1995). However, due to the relative faintness of this target, the 
N{\sc i} triplet was observed at medium resolution only, and other interstellar 
components cannot be ruled out as of now. Obviously this target looks very 
promising, providing these first observations be complemented with higher 
resolution, higher signal-to-noise ratio data.

In the case of G191--B2B, data were obtained in Cycles 1 and 5 at high
resolution ($\simeq$3.3 km s$^{-1}$) for Ly$\alpha$, N{\sc i} 1200\AA, 
O{\sc i} 1302\AA, Si{\sc ii} 1304\AA, Mg{\sc ii} 2800\AA, Fe{\sc ii}
2343\AA~and Si{\sc iii} 1206\AA. The line of sight velocity structure coherent 
in all these lines (about 15, including thoses observed at lower resolution) 
comprises one H{\sc i} region -- the Local Cloud observed toward 
Capella -- together with two H{\sc ii} regions; we refer to these components 
as blue, white and red, according to their positions along the wavelength scale.
Both H{\sc ii} regions are clearly seen in Si{\sc iii}, but their presence is 
also felt in strong lines such as H{\sc i} Ly$\alpha$ and O{\sc i} 1302\AA. The 
analysis in terms of column densities is still underway, but it seems already 
clear that the column density ratio of the blue to the red component varies 
from ion to ion. In particular, if the D/H ratio for the red component common 
to the G191--B2B and Capella sight-lines (these stars are separated by only 
$8^{o}$ on the sky) is forced to be that found by Linsky et al. (1993, 1995), 
then the D/H ratio for the blue component appears significantly lower.

In fig. 2 are superposed the G191--B2B spectra obtained at the highest 
resolution of Ech-A of the GHRS, all plotted in velocity space to allow an 
easy comparison of the different absorption features. In both N{\sc i} and 
Si{\sc ii} lines, at least two components are obviously present. However, only
one of them (at about +9 km s$^{-1}$, dotted line) is detected in the 
Si{\sc iii} line (the feature near +29 km s$^{-1}$~is stellar), while a third
component close to +13 km s$^{-1}$~is needed to fit the Si{\sc iii} line. The
existence of this third component is further confirmed in the O{\sc i} line in
order to deepen enough the O{\sc i} absorption feature between both N{\sc i} 
and Si{\sc ii} components. Freezing then the velociy separations and $b$-values,
it is possible to fit the D{\sc i} line. The stricking result is that the D/H 
ratio varies by at least a factor of three between the two extreme blue and red
components, which are better constrained. If one assumes that D/H is 
$1.6\times10^{-5}$ in the red component (which corresponds to the Capella one), 
a value certainly compatible with the total H{\sc i} content on that line of 
sight and with the known N{\sc i} abundance (Ferlet 1981), then the D/H ratio 
in the blue component has to be of the order of $5.5\times10^{-6}$. Furthermore,
if one considers that O{\sc i} is a better tracer of the H{\sc i} and D{\sc i}
gas than N{\sc i} because of its nearly identical ionization potential, then
the D/H ratio in the blue component should be even lower!

\begin{figure}[h]
\epsfysize=10cm
$$\epsfbox[18 144 592 718]{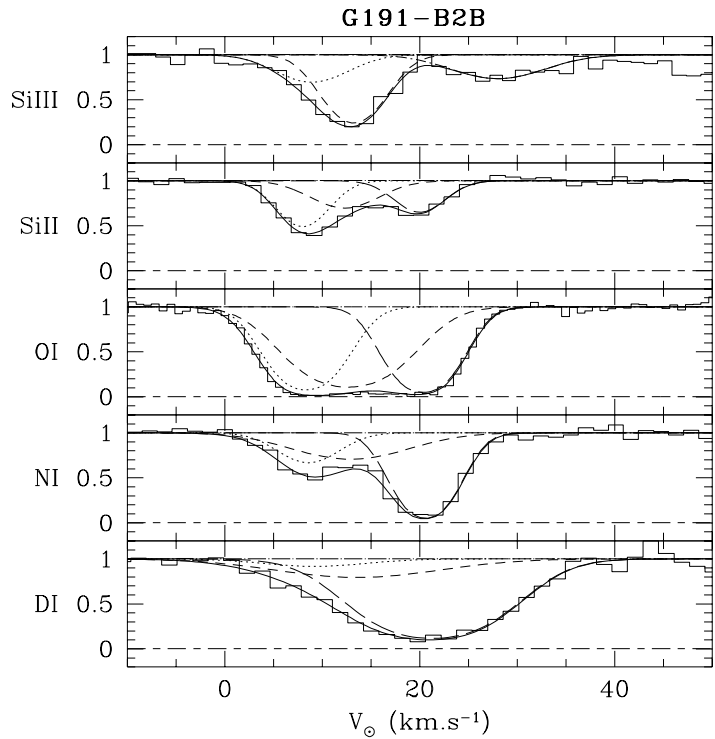}$$
\caption{Are superposed in velocity space the absorption features seen in the 
different species indicated on the left hand side of the figure. Two obvious 
components are seen in Si{\sc ii} and N{\sc i}. However, a third component in 
between the two others is imposed by the Si{\sc iii} and O{\sc i} lines. The 
two components at the shortest wavelengths, seen in Si{\sc iii}, are H{\sc ii} 
regions, while the component at $\sim$20 km s$^{-1}$ is the Local Cloud, an 
H{\sc i} region. It is clear that the D/H ratio cannot be the same in the red 
and the blue component (see text). Also, it is obvious that deducing lines of 
sight velocity structure for D/H evaluations from ionized species could be 
extremely misleading.}
\label{gbb}
\end{figure}

It is interesting to note that again D/H values clearly below 10$^{-5}$ seem 
to be required by the observations, suggesting again real D/H variations in
the local ISM. We have revealed a contamination of the H{\sc i} interstellar
absorption toward G191--B2B by residuals atoms of neutral hydrogen from two
H{\sc ii} regions. It is not unlikely that this effect is also present in other
lines of sight, in particular toward Capella. The observation of the whole
Lyman series, possible with the FUSE mission planned in 1998, will allow to 
better evaluate this effect. New HST--GHRS observations toward another white 
dwarf are presently underway.

\section{The D/H evaluations in the solar system}

The pre-solar value for the deuterium abundance has traditionally been derived
from meteorites and lunar soil: D/H=2.6$\pm1.0\times10^{-5}$~(Geiss 1993). 
The giant planets Jupiter and Saturn are considered to be undisturbed 
deuterium reservoirs, free from production or losses processes, preserving the 
abundance of their light elements since the formation of the solar system 4.5
billion years ago (Owen et al. 1986). The first measurements of the D/H ratio 
in the Jovian atmosphere have been performed through methane and its 
deuterated counterpart CH$_3$D yielding D/H=5.1$\pm2.2\times10^{-5}$~(Beer \& 
Taylor 1973). Abundant molecules like H$_2$ were also used yielding lower 
values: D/H=2.1$\pm0.4\times10^{-5}$~(Trauger et al. 1973). More recent
molecular measurements, also model-dependent, gives
D/H=2.5$\pm0.7\times10^{-5}$~(Gautier \& Owen 1983).

Recently, new important measurements of the D/H ratio implying very different 
methods were carried out. One is based on the far infrared ISO observation of 
the HD molecule in Jupiter (Encrenaz et al. 1996) and leads to
D/H=2.2$\pm0.5\times10^{-5}$, although some systematics are suspected. Another 
is based on the direct observation with HST--GHRS of both H{\sc i} and D{\sc i} 
Lyman $\alpha$~emission at the limb of Jupiter for the first time (Ben Jaffel 
et al. 1994, 1996). The last one is an in situ measurement with a mass 
spectrometer onboard the Galileo probe (Niemann et al. 1996). These two latter
yield similar ratios: 5.9$\pm1.4\times10^{-5}$~and
5.0$\pm2.0\times10^{-5}$~respectively.

These last values are higher than the meteoritic one (Geiss 1993) and the 
other Jovian ones (see e.g. Bjoraker et al. 1986) by $\sim2\sigma$. That may 
call to a reconsideration of the pre-solar abundance of deuterium since 
systematics may be inherent to each observational approach (see e.g. Lecluse 
et al. 1996). 

\section{Lines of sight toward QSOs}
\subsection{Observations}

A very promising approach to evaluate the D/H ratio was initiated a few years 
ago by observing directly with large ground based telescopes the Lyman lines 
redshifted in the visible part of the spectrum (see e.g. Carswell et al. 1987;
Webb et al. 1991). These lines are seen in absorption in QSOs spectra produced 
by the absorption of H{\sc i} gas present in the intergalactic medium in front 
of the QSOs. The most favorable case for detection of deuterium is provided in 
Lyman limit systems, whose H{\sc i} column densities typically range between
$\sim10^{16}-10^{19}$~cm$^{-2}$~and H{\sc i} $b$-values are $>15$ km s$^{-1}$. 
The mass and size of these clouds are essentially unknown by several orders of 
magnitudes.

The very interesting point of this approach is that one may probe directly
a primordial cloud ({\it i.e.} a very low metallicity system), and thus have
direct access to the primordial D/H value. On the other hand, the difficulty 
is twofold: {\it i)} the relative faintness of the sources renders these 
observations very difficult; {\it ii)} the number of absorbers per unit 
redshift and per unit column density increases with decreasing column density, 
so that there is always the possibility that the observed deuterium feature 
would be in fact a mimicking low-column density H{\sc i} absorber. The advent 
of very large telescopes can at least partly overcome the first point. 
Space-based UV telescopes, such as HST, offer the possibility of observing 
absorbers at low redshift, where the probability of contamination is greatly 
reduced. Note as well that there is a large scatter of metallicity with 
redshift, so that there are many low-redshift absorber candidates that are 
truly metal-poor (see Timmes et al. 1996). Thus, one can hope that a large 
number of measurements of the deuterium abundance in distant extragalactic 
objects, using both types of instrumentation, will yield its primordial 
abundance.

As for today, the different reported estimations are summarized in table 1.
For high redshift QSOs absorption systems, they are obtained from observations 
with the 10m Keck1 or the 4m Kitt Peak telescopes; for the lower redshift
system, data are from HST. Most were made with typical spectral resolutions of 
8 km s$^{-1}$ and signal-to-noise ratios at the D{\sc i} Ly$\alpha$ line which 
vary from $\sim10$ to as high as $\sim75$. Actually, only three detections of 
deuterium rather than upper limits are claimed, in complete disagreement with 
each other (1$\sigma$~random error in table 1), giving either a high 
($\sim2\times10^{-4}$, Rugers \& Hogan 1996a, 1996b) or a low 
($\sim2.5\times10^{-5}$, Tytler et al. 1996) D/H ratio.

\begin{table}
\caption{D/H measurements in absorption systems toward QSOs.}
\begin{center}\small
\begin{tabular}{l@{\hspace{1.5cm}}*{2}{c@{\hspace{1.5cm}}l}r}
QSO & z$_{abs}$ & D/H & Ref.\\
\hline
Q0014+8118 & 3.320 & $\leq2.5\times10^{-4}$ & 1,2\\
Q0014+8118 & 3.320 & $1.9\pm0.5\times10^{-4}$ & 3\\
Q1009+2956 & 2.504 & $2.5\pm0.5\times10^{-5}$ & 4\\
Q1937$-$1009 & 3.572 & $2.3\pm0.3\times10^{-5}$ & 5\\
Q1202$-$0725 & 4.672 & $\leq1.5\times10^{-4}$ & 6\\
Q0420$-$3851 & 3.086 & $>2.\times10^{-5}$ & 7\\
Q1937$-$1009 & 3.572 & $>4.\times10^{-5}$ & 8\\
Q0454$-$2203 & 0.482 & $<1.\times10^{-5}$ & 9\\
\hline
\multicolumn{4}{l}{Ref. 1 Songaila et al. 1994; 2 Carswell et al. 1994; 
3 Rugers \& Hogan 1996;}\\
\multicolumn{4}{@{\hspace{1cm}}l}{4 Burles \& Tytler 1996; 5 Tytler et al. 
1996; 6 Wampler et al. 1996;}\\
\multicolumn{4}{@{\hspace{1cm}}l}{7 Carswell et al. 1996; 8 Songaila et al. 
1996; 9 Webb et al. 1997.}
\end{tabular}
\end{center}
\end{table}

The generic method employed in these studies is as follows. Metal lines such as
C{\sc iv}, Si{\sc iv}, Mg{\sc ii} are used to determine the velocity structure 
in the redshift region where D{\sc i} is being chased. Taking this velocity 
structure into account, all the Lyman series up to the Lyman limit and the limit
itself are then fitted for estimating the H{\sc i} column density and 
$b$(H{\sc i}). In general, the high order lines yield an estimate of $b$, 
while the Lyman limit yields an estimate of N(H{\sc i}); if damping wings are 
present in Ly$\alpha$, they give also strong constraints on N(H{\sc i}). But 
one has to recall that again the usual limiting factor in determining the D/H 
ratio is the evaluation of N(H{\sc i}). Whenever D is detected at Ly$\alpha$, 
the estimate of D/H is rather secure if all the above steps have been 
successfully completed {\it i.e.} the velocity structure is in good agreement 
between the metal lines and the H{\sc i} lines, $b$ and N(H{\sc i}) have been 
accurately determined.

Although there is always the possibility that the D{\sc i} line be in fact an 
H{\sc i} interloper, the actual D{\sc i} detections were claimed on the basis 
of the following arguments. It is known (or believed) that the distribution of 
$b$-values in Lyman limit systems shows a cut-off around (and below) 
$b\sim15$ km s$^{-1}$. In the three claimed detections, the $b$-value for the 
D{\sc i} line was found in the order of~$\sim10$ km s$^{-1}$, with
$b$(D{\sc i})/$b$(H{\sc i})$\sim1/\sqrt{2}$~which corresponds to a pure thermal 
case. Moreover, the probability for interloping was estimated to be $<0.1$\%. 

Three caveats can at least be identified in this type of analysis, the 
magnitude of which are yet unknown. First, as already metionned, ionized metal 
species do not trace correctly the H{\sc i} gas. Hence, one should imagine 
that the redshift of H{\sc i} absorbers do not match those derived from metal 
lines (see a stricking example of such a situation in figure 3). Furthermore,
there could be substantial substructure in these H{\sc i} clouds, with no hope 
of tracing them from either H{\sc i} or ionized metal lines. An interesting 
study would be to observe these clouds at higher resolution ($<$ 8km s$^{-1}$) 
in O{\sc i} (N{\sc i} is not expected to be detectable in these very metal-poor 
clouds). Second, the estimate of N(H{\sc i}) from the Lyman limit is not 
reliable if the residual flux below the limit is uncertain, or is contaminated 
by instrumental noise (Songaila et al. 1996). Third, the asumption of Voigt 
profiles in the analysis could be extremely misleading as suggested by 
Levshakov \& Takahara (1996), particularly in the case of QSOs for which the
number of available absorption metallic lines is greatly reduced.

To increase the general confusion toward what should be considered as {\bf the}
primordial D/H value, a $2\sigma$ upper limit D/H$<1.\times10^{-5}$ has been
very recently set from HST observations of a z$_{abs}$=0.4823 absorber 
(Webb et al. 1997; table 1). These data were gathered in Cycles 5 and 6 at 
$\Delta\lambda\sim16$ km s$^{-1}$, and signal-to-noise ratios $\sim5-8$. They 
cover the Lyman limit, Ly$\epsilon$, Ly$\delta$, Ly$\gamma$, Ly$\alpha$, 
Si{\sc iii}; further ground-based observations of Mg{\sc ii} were collected at 
the Cerro Tololo International Observatory; see figure 3. The generic method 
outlined above was followed to derive D/H, although no Mg{\sc ii} is seen in
the system. Deuterium is not detected at Ly$\alpha$. But the position of the 
H{\sc i} absorber at Ly$\alpha$ is accurately known from the other lines, and 
precludes any deuterium absorption, whence the strong limit. This result could 
nevertheless be modified if substructure were present (this study is underway).

\begin{figure}[h]
\epsfysize=11cm
$$\epsfbox[18 144 592 718]{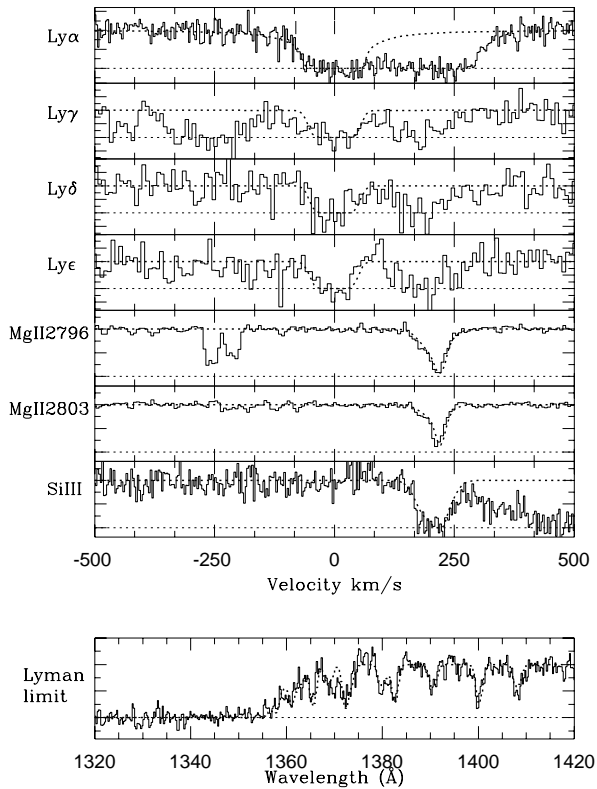}$$
\caption{Exemples of spectra of the z=0.482264, z=0.483036, and z=0.483304 
absorbers toward Q0454-2203, observed with HST--GHRS and FOS at Ly$\alpha$, 
Ly$\gamma$, Ly$\delta$, Ly$\epsilon$, Ly limit, and SiIII; the Mg{\sc ii} 
spectrum was collected at CTIO. The system at 0 km s$^{-1}$ is the one in 
which D/H has been evaluated to be $<1.\times10^{-5}$. Contrary to the other 
more complex system at about $+$200 km s$^{-1}$, it is not seen in either 
Mg{\sc ii} nor Si{\sc iii} and thus presents a low metallicity ($<$1/100 solar).}
\label{qso}
\end{figure}

\subsection{Who's wrong, who's right?}

If the result of Webb et al. (1997) is observationally confirmed toward some
other QSOs, then there will be definitely something rotten in some remote 
kingdom... No model of chemical evolution is able to account for an increase of
the deuterium abundance with metallicity. In that situation, we will thus have 
to assume either the distribution of cosmic deuterium is strongly inhomogeneous,
or such a low value is the result of a strong contamination by deuterium-poor 
gas. However, it is not so easy to deplete deuterium without producing metals, 
although a particular configuration of the line of sight aiming through a 
stellar wind contaminated region could do the job. A remote possibility is to 
produce significant amount of deuterium in some still unknown astrophysical 
site... Keeping these remarks in mind, let us discard this result from the 
following discussion, because too perturbing!

We now assume that Tytler et al. (1996) is right, {\it i.e.} the primordial 
deuterium abundance is low, $\sim2.5\times10^{-5}$~as prefered by Hata
et al. (1996). It is then easy to argue that the high deuterium abundances
sometimes measured are due to interlopers, eventhough their expected small 
probability of occurence, eventhough the claimed detected signature of D{\sc i}
$b$-value. In order to match a low primordial deuterium abundance with other
light elements, one would have to accomodate either large systematics in the 
determination of the primordial $^4$He mass fraction, or, equivalently, further
light degrees of freedom (equivalent to neutrino flavors) at BBN. The reader 
is refered to Copi et al. (1996a) for a discussion of this point. The ensuing 
chemical evolution of deuterium would be rather straightforward, possibly 
involving infall to avoid too much deuterium destruction.

Let now assume that Tytler et al. (1996) is wrong, {\it i.e.} the high values 
of D/H are the correct primordial ones. Although in excellent agreement with 
the primordial abundances of $^7$Li and $^4$He (Cardall \& Fuller 1996), such 
high values would lead to several other problems. The implied low baryonic
density parameter $\Omega_B$~would strengthen the so-called ``baryon cluster 
catastrophy''. The chemical evolution of deuterium would be quite difficult to 
account for, as it would require an astration factor $\sim15$~over $\sim10$Gyrs
when standard models destroy D by no more than a factor $\sim3$. Moreover, 
most models already overproduce $^3$He after 10Gyrs of evolution. Since D is 
converted to $^3$He in stellar interiors, a high D/H ratio would make this 
situation worse. However, Scully et al. (1996) have developped a promising new 
model which could resolve these problems.

Finally, let put everyone right and happy. The deuterium abundance seen in one 
absorber or the other, or both, could have been strongly affected by some 
unknown mechanism (e.g. Timmes et al 1996; Jedamzik \& Fuller 1996). However
there is, up to now, no compelling mechanism able to reconcile the observed 
values. As well and after all, there might not be a unique primordial deuterium
abundance. For instance, the presence of isocurvature baryon fluctuations at 
BBN would affect the yields differently in different regions. However, it seems
that Copi et al. (1996b) and Jedamzik \& Fuller (1996) do not quite agree on
wether or not such fluctuations of an amplitude and a scale large enough to 
explain the variations of D/H at high redshifts can be made compatible with 
the high degree of isotropy of the Cosmic Microwave Background (CMB).

Last, the remaining possibility is that no one is right (Levshakov \& Takahara
1996). For sure, that would be fun and will ask for many more observations! 

\section{Conclusion}

The previous discussion is very well summarized in figure 4. The different --
and discordant -- D/H evaluations are shown as a function of time, with no a 
priori bias to select one over another. Some of the differences are too large 
to be accounted for. It is clear that all measurements cannot be correct and 
some still unknown systematics should be identified. Certainly each one will 
find its preferred value in each domain, but our impression is that for the 
moment we are far from having understood the whole story.

\begin{figure}[h]
\epsfysize=9cm
$$\epsfbox[18 144 592 718]{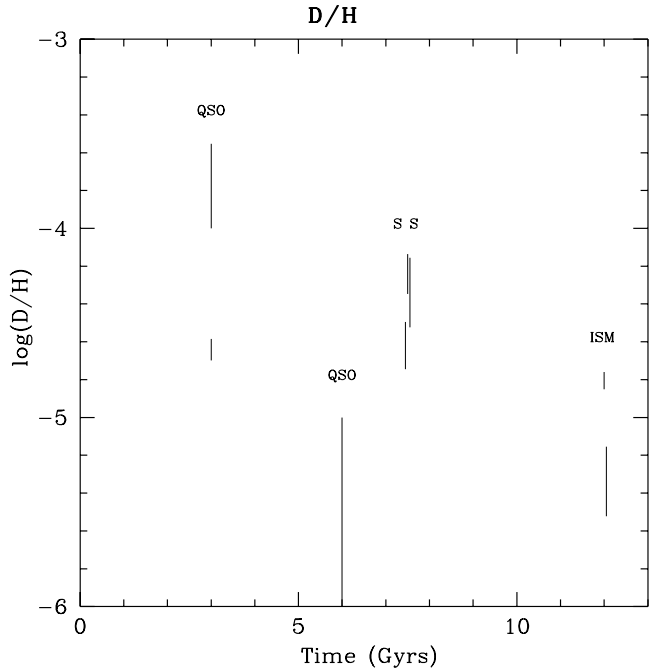}$$
\caption{The different D/H evaluations are shown approximately as a function 
of the epochs in the Universe when they are relevant. All the scatter is not 
shown but merely the main trends. For the ISM, the Capella value along with
possible significant low values; in the solar system, the standard and recent
values; for the QSOs, the most extreme cases at high redshifts and the only
measurement at lower redshift.}
\label{conc}
\end{figure}

As a matter of fact, if the variations of the D/H ratio in the local 
interstellar medium are illusory, then one could quote as an average of the 
published values: (D/H)$_{ISM}\simeq1.3\pm0.4\times10^{-5}$. The rather large 
error bar arises from a subjective although conservative viewpoint. On the 
contrary, if D/H does vary in the ISM, one has to understand why; until then, 
{\it no measurement of the D/H ratio in the ISM or the IGM should be quoted as 
reliable}. Moreover, one should expect these variations to be larger in 
reality. The actual value might in fact be very different from what is 
observed if these variations are systematic, {\it i.e.} act in one way only; 
this in turn would heavily bear on the chemical evolution of deuterium. It 
also appears that the upper bound on $\Omega_B$~is obtained from BBN 
predictions through the interstellar abundance of deuterium: this bound would 
have to be removed until the variations and their cause are properly understood.

There is a hope that the FUSE mission will solve these problems from 1998. It 
will probe the ISM further than the local medium, up to extragalactic 
low-redshift objects. It will look for gradients of the deuterium abundance 
with galactocentric distance and with galactic height in the halo. These 
dedicated studies with FUSE should greatly clarify the problem of the chemical 
evolution of deuterium. 

In other words, when some important observable is evaluated:

\medskip
\centerline{One measurement is nice}
\medskip
\centerline{Few measurements are a crisis}
\medskip
\centerline{Many more measurements could be fun}
\vskip 1cm

\noindent
{\bf Acknowledgments.}
The study of the interstellar D/H ratio toward white dwarves with HST has been
initiated by A.V-M and R.F. Subsequently, several French and US collaborators 
were included. M.L. acknowledges support by the NASA, DoE, and NSF at the 
University of Chicago. We thank C. Copi, D. Schramm, J. Truran, M. Turner and 
D. York for useful discussions.
\vskip 1cm

\reference Audouze, J., \& Tinsley, B.M. 1976, ARAA,
    14, 43

\reference Allen, M., Jenkins, E.B., \& Snow T.P.  1992, \apj S, 83, 261

\reference Beer, R., \& Taylor, F. 1973, \apj, 179, 309

\reference Ben Jaffel, L. et al. 1994, BAAS, 26, 1100

\reference Ben Jaffel, L. et al. 1996, these proceedings

\reference Bjoraker, G., Larson, H., \& Kunde, V. 1986, Icarus, 66, 579

\reference Boesgaard, A.M., \& Steigman, G. 1985, ARAA,
    23, 319

\reference Bruston, P., Audouze, J., Vidal-Madjar, A., \& Laurent, C.  1981, 
    \apj, 243, 161

\reference Burles, S., \& Tytler, D. 1996, submitted to Science, preprint
    {\tt astro-ph/9603070}

\reference Cardall, G.M., \& Fuller, G. 1996, submitted to \apj, preprint 
    {\tt astro-ph/9603071}

\reference Carswell, R.F., Webb, J.K., Baldwin, J.A., \& Atwood, B. 1987, \apj,
    319, 709

\reference Carswell, R.F. et al. 1994, \mnras, 268, L1

\reference Carswell, R.F. et al. 1996, \mnras, 278, 506

\reference Copi, C.J., Schramm, D.N., \& Turner, M.S. 1996a, submitted to 
    Phys. Rev. Lett., preprint {\tt astro-ph/9606059}

\reference Copi, C.J., Olive, K.A., \& Schramm, D.N. 1996b, submitted to \apj,
    preprint {\tt astro-ph/9606156}

\reference Dring, A., Murthy, J., Henry, R.C. et al. 1997, in preparation

\reference Encrenaz, T. et al. 1996, \astap, in press

\reference Ferlet, R. 1981, \astap, 98, L1

\reference Ferlet, R. 1992, in IAU Symposium 150, 85

\reference Gautier, D., \& Owen, T. 1983, Nature, 304, 691

\reference Geiss, J. 1993, in Origin and Evolution of the Elements, CUP, 89

\reference Geiss, J., \& Reeves, H. 1972, \astap, 18, 126

\reference Gry, C., Laurent, C., \& Vidal-Madjar, A. 1983, \astap, 124, 99

\reference Gry, C. et al. 1995, \astap, 302, 497

\reference Hata, N., Steigman, G., Bludman, S., \& Langacker, P. 1996,
    submitted to \apj, preprint {\tt astro-ph/9603087}

\reference Heiles, C., McCullough, P., \& Glassgold, A. 1993, \apj S, 89, 
    271

\reference Jedamzik, K., \& Fuller, G. 1995, \apj, 452, 33

\reference Jedamzik, K., \& Fuller, G. 1996, submitted to \apj, preprint
    {\tt astro-ph/9609103}

\reference Jenkins, E.B. 1996, submitted to \apj, preprint

\reference Lecluse, C., Robert, F., Gautier, D., \& Guiraud, M. 1996, 
    Plan. Space Sci., in press

\reference Levshakov, S.A., \& Takahara, F. 1996, \mnras, 279, 651

\reference Levshakov, S.A., Kegel, W.H., \& Mazets, I.E. 1996, Yukawa Institut,
    Kyoto preprint YITP-96-23 

\reference Landsman, W., Sofia, U.J., \& Bergeron, P. 1996, in Science with 
    the Hubble Space Telescope - II, STScI, 454

\reference Laurent, C., Vidal-Madjar, A., \& York, D.G. 1979, \apj, 229, 923

\reference Lemoine, M. et al. 1996, \astap, 308, 601

\reference Linsky, J. et al. 1993, \apj, 402, 694

\reference Linsky, J. et al. 1995, \apj, 451, 335

\reference Linsky, J., \& Wood, B.E. 1996, \apj, 463, 254

\reference Niemann, H.B. et al. 1996, Science, 272, 846

\reference Olive, K., Schramm, D., Steigman, G., \& Walker, T. 1990, Phys. 
    Lett., B236, 454

\reference Owen, T., Lutz, B., \& De Bergh, C. 1986, Nature, 320, 244

\reference Pagel, B. et al. 1992, \mnras, 255, 325

\reference Piskunov, N. et al. 1996, \apj, in press

\reference Prantzos, N. 1996, \astap, 310, 106

\reference Rogerson, J., \& York, D. 1973, \apj, 186, L95

\reference Rugers M., \& Hogan, C.J. 1996a, \apj, 549, L1

\reference Rugers M., \& Hogan, C.J. 1996b, AJ, 111, 2135

\reference Scully, S. et al. 1996, preprint {\tt astro-ph/9607106}

\reference Songaila, A., Cowie, L.L., Hogan, C.J., \& Rugers, M. 1994, Nature,
    368, 599

\reference Songaila, A., Wampler, E.J., \& Cowie, L.L. 1996, submitted to 
    Nature, preprint

\reference Timmes, F.X., Truran, J.W., Lauroesch, J.T., \& York, D.G. 1996,
    submitted to \apj, preprint

\reference Trauger, J., Roesler, F., Carleton, N., \& Traub, W. 1973, \apj,
    184, L137

\reference Tytler, D., Fan, X.-M., \& Burles, S. 1996, Nature, 381, 207 

\reference Vangioni-Flam, E., \& Cass\'e, M. 1994, \apj, 427, 618

\reference Vidal-Madjar, A., Laurent, C., Bruston, P., \& Audouze, J.  1978, 
    \apj, 223, 589

\reference Vidal-Madjar, A., \& Gry, C. 1984, \astap, 138, 285

\reference Vidal-Madjar, A.  1991, Adv. Space Res., 11, 97

\reference Vidal-Madjar, A., Lemoine, M., Ferlet, R., \& Gry, C. 1997,
    in preparation

\reference Wampler et al. 1996, \astap, in press 

\reference Webb, J.K. et al. 1991, \mnras, 250, 657

\reference Webb, J.K. et al. 1997, in preparation

\reference York, D.G. 1983, \apj, 264, 172

\reference York, D.G. et al. 1983, \apj, 266, L55

\end{document}